\documentclass[12pt]{article}
\usepackage[T1]{fontenc}
\usepackage{geometry}
\usepackage{palatino}
\usepackage{mathpazo}
\usepackage{mathtools}
\usepackage{amssymb}
\usepackage[nolist]{acronym}
\usepackage{dsfont}
\usepackage[small]{caption}
\usepackage{mathrsfs}	
\usepackage{cite}
\usepackage{cancel}
\usepackage{nicefrac}
\usepackage{pifont}
\usepackage{tikz}
\definecolor{NavyBlue}{rgb}{0.0, 0.0, 0.5}
\usepackage[linkcolor=blue,urlcolor=blue,citecolor=NavyBlue,colorlinks=true]{hyperref} 
\hypersetup{linktocpage}
\usepackage{eqparbox}
\usepackage[boxsize=6 pt, centertableaux]{ytableau}
\usepackage{young}
\usepackage{braket}
\usepackage{xfrac}
\usepackage{xspace}
\usepackage{booktabs}
\usepackage{mleftright} \mleftright
\usepackage{subcaption}
\usepackage{letltxmacro}

\usepackage[textsize=tiny]{todonotes} \usepackage{soul} \usepackage{cleveref}

\makeatletter \g@addto@macro\bfseries{\boldmath} \makeatother

\DeclareMathOperator{\Pf}{Pf}

\newcommand{\SO}[1]{\ensuremath{\mathrm{SO}(#1)}}
\newcommand{\SU}[1]{\ensuremath{\mathrm{SU}(#1)}}
\newcommand{\SP}[1]{\ensuremath{\mathrm{SP}(#1)}}

\newcommand{\U}[1]{\ensuremath{\mathrm{U}(#1)}}

\newcommand{\x}{\ifmmode\times\else\ensuremath{\times}\fi}

\newcommand{\rep}[1]{\ensuremath{{\boldsymbol{#1}}}}

\def\barM{\ensuremath{\overline{M}}}
\def\barQ{\ensuremath{\overline{Q}}}
\def\barA{\ensuremath{\overline{A}}}
\def\cA{\ensuremath{\mathcal{A}}}
\def\barcA{\ensuremath{\overline{\mathcal{A}}}}
\def\barS{\ensuremath{\overline{S}}}
\def\cS{\ensuremath{\mathcal{S}}}
\def\barcS{\ensuremath{\overline{\mathcal{S}}}}
\def\cM{\ensuremath{\mathcal{M}}}

\def\cB{\ensuremath{\mathcal{B}}}
\def\barcB{\ensuremath{\overline{\mathcal{B}}}}
\def\cQ{\ensuremath{\mathcal{Q}}}

\newcommand{\sconf}[0]{$s$-confining\xspace}

\newcommand{\sconfinement}[0]{$s$-confinement\xspace}

\newcommand{\SCONF}{$s$-confinement\xspace}

\newcommand{\SUTS}{\ifmmode\SU2_s\else$\SU2_s$\xspace\fi}
\LetLtxMacro\OldDot\.
\renewcommand{\.}{\ifmmode\hspace{0.1ex}\else\OldDot\fi}
\LetLtxMacro\OldOverline\overline
\renewcommand{\overline}[1]{{\hspace{0.1ex}\OldOverline{\hspace{-0.1ex}#1\hspace{-0.033ex}}\hspace{0.033ex}}}

\allowdisplaybreaks

\begin{acronym}
  \acro{QFT}{Quantum Field Theory}
  \acrodefplural{QFT}{Quantum Field Theories}
  \acro{SYM}{super Yang-Mills}
  \acro{GUT}{Grand Unified Theory}
  \acro{IR}{infrared}
  \acro{NS}{Nelson and Strassler}
  \acro{RG}{renormalization group}
  \acro{NS}{Nelson--Strassler}
  \acro{RT}{Razamat--Tong}
  \acro{SM}{Standard Model}
  \acro{SUSY}{supersymmetric}
  \acro{MSSM}{minimal supersymmetric Standard Model}
  \acro{SYM}{super-Yang--Mills}
  \acro{VEV}{vacuum expectation value}
  \acrodefplural{VEV}{vacuum expectation values}
  \acro{UV}{ultraviolet}
\end{acronym}

\begin{document}
  \newcommand\mytitle{Chirality Changing RG Flows: Dynamics and Models}
  \newcommand\mypreprint{UCI-TR-2023-01}
  \begin{titlepage}
    \begin{flushright} \mypreprint \end{flushright}
		
    \vspace*{2cm}
		
    \begin{center} {\Large\sffamily\bfseries\mytitle}
			
    \vspace{1cm}
			
    \renewcommand*{\thefootnote}{\fnsymbol{footnote}}
			
    \textbf{%
      Yuri~Shirman$^{a,}$\footnote{yshirman@uci.edu},
      Shreya~Shukla$^{a,}$\footnote{sshukla4@uci.edu} and   Michael~Waterbury$^{b,}$\footnote{mwaterbu@uci.edu} }
      \\[8mm] \textit{$^a$\small
        ~Department of Physics and Astronomy, University of California, Irvine, CA 
        92697-4575, USA }
      \\[5mm] \textit{$^b$~\small Physics Department, Technion - Israel Institute of Technology, Technion city, Haifa 3200003, Israel}

    \end{center}
		
    \vspace*{1cm}
    \begin{abstract}
    Chirality plays an important role in understanding the dynamics of quantum field theories. In this paper, we study the dynamics of models where renormalization group flows change the chiral structure of the theory. We introduce model building tools and construct models with a variety of chirality flows: from the appearance of new massless composite matter, to the development of mass gaps to completely general changes in the chiral matter content. The stability of chirally symmetric vacua is sensitive to the interplay between non-perturbative dynamics and deformations necessary to generate chirality flows. In particular, we show that chirality flows can be easily induced by deformations of \sconf models. On the other hand, in the absence of true \sconfinement, the required  deformations destabilize chirally symmetric ground states.
    \end{abstract}

  \vspace*{1cm}

  \end{titlepage}
	
  \noindent\rule{\linewidth}{0.4pt}
  \tableofcontents
  \noindent\rule{\linewidth}{0.4pt}
	
\setcounter{footnote}{0} 
  \section{Introduction} \label{sec:Introduction}

  Chiral symmetries play an essential role in studying the dynamics of
  \acp{QFT}. Since mass terms break chiral symmetries, they are only allowed
  for fermions in vectorlike representations, while fermions in theories with
  chiral matter content must remain massless unless chiral symmetries are
  broken spontaneously. It seems almost obvious that these
  statements are \ac{RG} invariant. However, examples of \ac{RG} flows altering
  the chiral structure of \acp{QFT}  have been known for some time
  \cite{Strassler:1995ia,Nelson:1996km,Razamat:2020kyf,Tong:2021phe}. The
  underlying physics relies on the existence of models exhibiting confinement
  without chiral symmetry breaking \cite{Seiberg:1994bz} referred to as s-confinement~\footnote{See
  \cite{Csaki:1996zb} for a complete classification of s-confining theories.}. Generically, the elementary and low--energy degrees of
  freedom of s-confining theories transform in different representations of the
  global symmetries. Thus the chiral structure of the matter sector may differ
  between \ac{UV} and \ac{IR}. The early models of this type
  \cite{Strassler:1995ia,Nelson:1996km} were motivated by a search for
  realistic \ac{SUSY} extensions of the \ac{SM} and contained composite
  massless \ac{SM} generations in the \ac{IR}. More recently, the possibility of
  developing a mass gap in theories with apparently chiral matter content
  attracted some attention.
   References \cite{Razamat:2020kyf,Tong:2021phe} explored the
    deformation class of \acp{QFT} by constructing flows in theory space from
    anomaly-free chiral theories to the trivial theory with no massless
    fermions. The ideas introduced in  \cite{Strassler:1995ia,Nelson:1996km, Razamat:2020kyf,Tong:2021phe} were used in \cite{Ramos-Sanchez:2021woq} 
    to argue that string compactifications may lead to realistic low energy physics even if the number of chiral generations in the UV differs from $3$. The authors of \cite{Ramos-Sanchez:2021woq} also began a careful analysis of dynamics underlying the chirality changing \ac{RG} flows.  The goal of this paper is to complete the systematic analysis of this phenomenon and  elucidate a unifying picture of chirality
  changing \ac{RG} flows. While we will concentrate on the generation of mass
  gap, our analysis will also cover models where additional composite chiral
  multiplets appear in the \ac{IR} as well as more general cases where the
  chiral matter content in \ac{IR} differs from that in \ac{UV}. 
  
  The model-building prescription for generating a mass gap is quite simple: one deforms an \sconf gauge theory \cite{Csaki:1996zb} by introducing the superpotential couplings to a set of spectators superfields, transforming under the chiral symmetry in representations conjugate to the representations of the composites of the strong dynamics. The most general superpotential allowed by such a deformation of the \sconf model lifts all classical flat directions of the \sconf sector
  ensuring that
  in the ground state the strong \sconf group is unbroken and confines. One
  must then verify that the classical flat directions associated with the
  spectator superfields are also lifted, which will generically be the case. If
  the spectator flat directions are indeed lifted, the global symmetry group of
  the \sconf sector is unbroken and a chiral subgroup of the global symmetry may be gauged thus
  leading us to the desired result. On the other hand, we will also see examples where the flat directions associated with the spectator fields are destabilized by the non-perturbative dynamics. Since the spectators are charged under the chiral sector the chiral symmetry is broken in this class of models. Finally, if one is interested in the appearance of chiral composite generations, one chooses different representations for the spectators so that in the \ac{IR} some or all of the composites do not have partners to generate mass terms.

  The paper is organized as follows. In \cref{sec:construct}, we discuss the
  general construction in more detail and explain the role of the interplay between tree level superpotential and non-perturbative \sconf dynamics in the stabilization of chirally symmetric vacua. In \cref{sec:SP}, we construct
  strongly-coupled \SP{2N} models which gap chiral matter containing an
  antisymmetric. We also show that models with dynamically generated mass gaps and composite chiral matter represent two examples of the same phenomenon. In \cref{sec:SO}, we explore an \SO{N} model where naively
 one expects a dynamically generated mass gap for chiral matter in a symmetric representation and show how this
  model fails. In \cref{sec:SU}, we examine a rich space of chirality-changing \ac{RG} flows in models based on the strong \SU{N} dynamics. Within this class of
  models, we construct a model which gaps symmetric matter content and
  illustrate how to generalize the construction to gap arbitrary
  representations.  We make concluding remarks in  \cref{sec:conc}.
  
  \section{Generic Construction} \label{sec:construct}
	
To construct models of chirality changing \ac{RG} flows we will adopt the model-building approach of 
  \cite{Razamat:2020kyf,Tong:2021phe} taking a product
  group theory $H \times G$ as a starting point. Here $G$ is the chiral symmetry group of interest which may be either a weakly-coupled gauge group or an
  anomaly-free global symmetry, while $H$ is the gauge group of an \sconf sector whose dynamics is responsible for the chirality flows. For now, assume that $G$ is unbroken by the confining dynamics of $H$, such that it is sensible to study the chiral properties of $G$
  in both the \ac{UV} and \ac{IR}. Fields charged under both $G$
  and $H$ confine into composites which generically transform under
  tensor representations of $G$ and have different chiral properties than the
  elementary representations of $G$. We call these flows from the \ac{UV} to
  \ac{IR} chirality changing flows on $G$ induced by $H$. We are particularly interested in deformations of \sconf models where some or all of the composite fields pair up with the spectators of the strong dynamics in vector-like representations.
  When this is the case, the vector-like representations can
  be decoupled with the addition of superpotential interactions that may be marginal or irrelevant in the \ac{UV} but behave as mass terms in the \ac{IR}.
  As we will show, the fact that the \ac{IR} mass terms originate from the dimension $d>2$ operators in the \ac{UV} implies that dynamical effects of these superpotential terms are quite non-trivial and may disrupt the confining dynamics of $H$. 
We will draw
  special attention to these scenarios. 

In this paper, we  restrict our attention to   $H \times G$ models with $\mathcal N=1$ \ac{SUSY}.  In our analysis we will be able to employ familiar tools often used in the study of dynamical supersymmetry breaking \cite{Shirman:1996jx} even though  the models we consider
  will possess supersymmetric ground states.

  Let us discuss the construction in a bit more detail. We will start with
  \sconf models based on gauge group $H$ and matter fields $Q_i$ transforming in
  a chiral representation of $H\times\widetilde{G}$, where $\widetilde{G}$ is a
  possibly anomalous chiral symmetry of the theory. We will limit our
  attention to an anomaly-free subgroup $G$ of this global symmetry and thus
  will study $H\times G$. As long as $G$ is only a global symmetry, the
  anomaly freedom means that mixed $H^2G$ anomalies cancel. The anomaly
  cancellation condition is automatically satisfied whenever $G$ is non-Abelian, continuous,
  and only imposes nontrivial constraints on the model when $G$ contains
  $\U{1}$ factors.
      Aside from these weak constraints, $G$ could be identified with any subgroup of
      $\widetilde{G}$. Generically $G$ will have cubic anomalies. These are
      harmless as long as $G$ is a global symmetry, however, we will imagine
      weakly gauging $G$. This is only possible if we add a set of spectators
      charged under $G$ whose contribution to cubic anomaly cancels the
      contribution of $Q_i$'s. 
  The dynamics of our \sconf model can be described in terms of the gauge invariant composites $\cM_f$.
  In the UV these composites scale as $\cM_f\sim Q_i^{d_f}$ and thus have
  engineering mass dimension $d_f$.%
  \footnote{Here our notation for the composites
    derives from the simplest case of a bilinear composite, a meson, $\cM\sim Q^2$. We
    stress that in this general discussion, $\cM_f$ represent all moduli of the
    theory regardless of their engineering dimension.}
  In the \ac{IR} the composite moduli $\cM_f$ are weakly coupled and have mass
  dimension one. Generically, $\cM_f$ will transform in chiral representations of
  $G$ and will contribute to cubic anomalies of $G$. The 't Hooft anomaly
  matching condition ensures that the $\cM_f$ saturate the anomalies of the
  microscopic theory. To be able to gauge $G$ we must introduce a set of spectator fields that cancel $G^3$ anomalies. The choice of spectators is not unique. For example, one can choose spectators $\overline{q}_i$ to transform in representations conjugate to those of elementary fields, $Q_i$, or a different set of spectators $\overline{M}_f$ transforming in representations conjugate to the composites $\cM_f$. As we shall soon see, the former choice may lead to an appearance of massless chiral composites of $G$ in the \ac{IR} while the latter choice may allow an \ac{RG} flow to a gapped vacuum.
 
  For the moment we choose the spectators transforming as $\overline{M}_f$ so that an \ac{IR} mass term is allowed in the superpotential
  \begin{equation}
    \label{eq:W-IR} \mathscr{W}=\sum_f \cM_f\overline{M}_f\,.
  \end{equation}
  We must remember that $H$ is \sconf, thus the full non-perturbative
  superpotential takes the form
  \begin{equation}
    \mathscr{W}=f(\cM_f,\Lambda)+\sum_f \cM_f\overline{M}_f\,,
  \end{equation}
  where $f(\cM_f,\Lambda)$ is a dynamical superpotential generated by the \sconf
  dynamics of $H$.
	
 By construction the full $H\times G$ symmetry is chiral and the mass terms are not allowed. Moreover, even the $G$ sector alone is chiral in the \ac{UV}. In the \ac{IR} the strongly coupled $H$ sector confines while the low energy matter content is vector-like under $G$. As long as the deformation (\ref{eq:W-IR}) of the \sconf model does not lift the chirally symmetric vacuum at the origin of the moduli space, the dynamics of the deformed \sconf theory results in  development of the mass gap in the \ac{IR}.
 
For chiral symmetry to be unbroken in the \ac{IR}, the \acp{VEV} of both the composite moduli $\cM_f$ and the spectators $\overline{M}_f$ must vanish in the ground state. This is indeed true for the moduli $\cM_f$ since the deformation  (\ref{eq:W-IR}) lifts all classical flat directions of $H$ as long as it contains mass terms for all $H$ moduli. However, while the deformation (\ref{eq:W-IR}), when written in terms of the \ac{IR} degrees of freedom, looks like a simple set of mass terms for all the spectators, the interplay between the non-perturbative dynamics of the \sconf sector and the tree level superpotential is quite non-trivial and may result in the spontaneous breaking of $G$. Indeed, while (\ref{eq:W-IR}) lifts all the classical flat directions of $H$, it introduces new classical flat directions parameterized by $\overline{M}_f$. To see that one simply needs to look at the deformation in terms of elementary degrees of freedom
  \begin{equation} \label{eq:W-UV}
    \mathscr{W}=\sum_f \left(Q_i\right)^{d_f} \overline{M}_f\,.
  \end{equation}
The extrema of this superpotential with respect to $\overline{M}_f$ are found at $\cM_f=0$ or, equivalently at $Q_i=0$. On the other hand, the extrema with respect to $Q_i$ are given by
  \begin{equation}
    \sum_f \frac{\partial 
    \cM_f}{\partial Q_i} \overline{M}_f=0\,,
  \end{equation}
  which is satisfied for all values of $\overline{M}_f$ since in the UV the
  composites $M_f$ are simply monomials of $Q_i$'s with dimensions greater than or
  equal to two.
  
As we will see in the following sections the interplay between the strong dynamics and the deformation (\ref{eq:W-IR}) generates a non-perturbative superpotential for the spectators. This is most easily seen by considering physics along classical flat directions for spectators that couple to mesons of strong dynamics. Along such flat directions the spectator \acp{VEV} generate large masses for all the quarks $Q_i$ and the low energy physics is described in terms of a pure \ac{SYM} theory with dynamical superpotential generated by gaugino condensation:
  \begin{equation}
    \mathscr{W}=\Lambda_L^3= \left(\overline{M}^{F}\Lambda^{b}\right)^{3/b_L}\,,
  \end{equation}
  where $\Lambda_L$ is the dynamical scale of the low energy SYM theory, $b$ and $b_L$ are one-loop
  beta-function coefficients of the UV and IR theories respectively, $F$ is the
  effective number of flavors in our \sconf UV model and in
  the second equality we used the scale matching relation
  $\Lambda_L^{b_L}=\overline{M}^F\Lambda^b$. As long as $3F/b_L>1$ the dynamical superpotential stabilizes the spectators near the origin, the analysis of the ground state in terms of the IR degrees of freedom is valid and the mass gap is generated. This is the case, for example, in models satisfying the \sconfinement conditions \cite{Seiberg:1994bz,Csaki:1996zb}.
  On the other hand, whenever $3F/b_L<1$ the dynamical superpotential destabilizes the chirally symmetric vacuum near the origin and the models of this type cannot lead to a mass gap.

  Before moving on to the examples, let us make a connection to
  \ac{RT}~\cite{Razamat:2020kyf} language. The discussion of \cite{Razamat:2020kyf} takes the model with chiral symmetry group $G$ as a starting point, then assigns some, but not all, chiral superfields charges under the strongly coupled $H$ sector. In this language, the spectators $\overline{M}$ represent the basic chiral matter of the \ac{UV} description.
  This is in contrast to our construction where $\overline{M}$ fields
  are spectators needed to generate a mass gap in an \sconf model. Nevertheless,
  once a model is fully specified we achieve the same result as in \cite{Razamat:2020kyf} --- a
  chiral theory with a mass gap in the \ac{IR}.

  \section{Chirality flows and $\SP{2N}$ dynamics} \label{sec:SP}
  In this section we consider the simplest class of models exhibiting chirality flows. These models are based on the \sconf models with $\SP{2N}$ gauge group with $F=N+2$  chiral matter fields in the fundamental representation. We will identify the chiral symmetry group $G$  with a subgroup of $\SU{2F}$, the maximal chiral symmetry of the $\SP{2N}$ dynamics. In section \ref{asym-even} we consider an example of a dynamically generated mass gap \cite{Razamat:2020kyf}, while studying a closely related example of a composite massless generation \cite{Nelson:1996km} in section \ref{sec:composite-gen}. In section \ref{sec:otherembed} we briefly discuss additional chirality flow models that can be obtained by considering different embeddings of $G$ into the maximal global symmetry of the \sconf sector.
  
  \subsection{Dynamically generated mass gap}
\label{asym-even}
Following  \cite{Razamat:2020kyf} we consider $\SP{2N}$ models where the chiral symmetry group $G$ is identified with the maximal global symmetry of the \sconf sector, $G=\SU{2F}=\SU{2N+4}$. To analyze the non-perturbative dynamics of this class of models we recall that an $\SP{2N}$ theory with $F$ flavors has an $\SU{2F}$ global symmetry and posseses a set of classical flat directions \cite{Intriligator:1995ne} which, up to gauge and global symmetry transformations,  can be parameterized by\footnote{Here we have restricted our attention to the $F>N$ case.}
\begin{equation}
Q=\left(\begin{array}{cccc} q_1&&&\\&q_2&&\\&&\ldots&\\&&&q_F\\&&&\\&&&\end{array} \right)\otimes\mathbb{1}_2\,.
\end{equation}
Alternatively, the space of classical vacua can be parameterized in terms of mesons, $\cA_{ij} \sim Q_iQ_{j}$ transforming in an antisymmetric representation of the global $\SU{2F}$ symmetry. At a generic point on the moduli space $\mathrm{rank}(\cA)=\mathrm{min}(2N,2F)$. This means that for $F>N$ the mesons must satisfy a set of constraints. Specifically in the case of interest, $F=N+2$, the meson \acp{VEV} satisfy  classical constraints
\begin{equation}
 \epsilon^{i_1\ldots i_{2N+2}}\cA_{i_1i_2}\cA_{i_3i_4}\ldots \cA_{i_{2N+1}i_{2N}}=0\,.
\end{equation}
These constraints may be compactly written as 
\begin{equation}
 \frac{\partial}{\partial \cA }\left(\mathrm{Pf}\, \cA\right)=0\,.
\end{equation}
Following Seiberg's analysis \cite{Seiberg:1994bz}  of \SCONF in $\SU{N}$, Intriligator and Pouliot argued that the quantum and classical space coincide in $F=N+2$ $\SP{2N}$ models. Since the origin belongs to the quantum moduli space, the model posseses a supersymmetric vacuum with unbroken chiral symmetry. The low energy physics is described in terms of mesons with a non-perturbative superpotential~\footnote{The equations of motion following from this superpotential enforce classical constraints on mesons $\cA$.} 
\begin{equation}
\label{eq:SP-dyn}
 W_\mathrm{dyn}=\frac{1}{\Lambda^{2N+2}}\mathrm{Pf}\cA\,\,.~
\end{equation}

To generate the mass gap \cite{Razamat:2020kyf} we deform the theory by including a set of spectator superfields $\barA$ transforming in the conjugate antisymmetric representation of the chiral $\SU{2F}=\SU{2N+4}$ symmetry with the tree superpotential 
\begin{equation}
\label{eq:SP-deform}
 W_\mathrm{tree}=\barA Q^2\sim \Lambda \barA \cA\,,
\end{equation}
where the second expression is written in terms of mesons $\cA$.
The \ac{UV} and \ac{IR} matter content of the model is presented in the top and bottom parts of table \ref{tab:SPN} respectively:
 \begin{table}[h!] \centering
      \begin{tabular}{cccc}
  \toprule & \SP{2N} & $\SU{2N+4}$ & $\U{1}_R$\\ \midrule $Q$ & $\ytableaushort{~}$ &
    $\ytableaushort{~}$ & $\frac{1}{N+2}$ \\
  $\barA$ & $\boldsymbol{1}$ & $\overline{\ytableaushort{\\}}\vphantom{\overline{\ytableaushort{\\}}^2}$ & $\frac{2N+2}{N+2}$ \\
  \midrule $\cA\sim Q^2$ & $\boldsymbol{1}$ & $\ytableaushort{\\}$ & $\frac{2}{N+2}$ \\
  $\barA$ & $\boldsymbol{1}$ & $\overline{\ytableaushort{\\}}\vphantom{\overline{\ytableaushort{\\}}^2}$ & $\frac{2N+2}{N+2}$ \\
  \bottomrule
    \end{tabular}
    \caption{Field content of the $\SP{2N}$ model with $F=N+2$ flavors.}
    \label{tab:SPN}
  \end{table}
 
 The \ac{IR} form of the superpotential (\ref{eq:SP-deform}) suggests that all the fields in the low energy effective theory become massive and the model possesses a unique vacuum at the origin with an unbroken chiral symmetry. While ultimately correct in this model, the conclusion requires a more careful analysis of the non-perturbative dynamics. Indeed, while the tree level superpotential lifts all flat directions associated with  $\SP{2N}$ gauge group, the deformed theory has a new set of classical flat directions parameterized by spectators $\barA$. Far enough along this new branch of classical vacua, $\barA\gg\Lambda$, the theory is weakly coupled and the analysis of dynamics is most easily performed in terms of quark superfields since their K\"ahler potential is nearly canonical in this regime. The spectator \acp{VEV} generate mass terms for quark superfields which can be integrated out. The low energy physics is then described as a pure \ac{SYM} theory whose coupling constant is field dependent:
\begin{equation}
 \Lambda_L^{3(N+1)}=\Pf \barA \Lambda^{2N+1}\,.
\end{equation}
In the \ac{IR} pure \ac{SYM} dynamics generates the gaugino condensate superpotential, which can also be interpreted as a superpotential for the spectators
  \begin{equation}
    W=\Lambda_L^3=\left(\mathrm{Pf}\left(\barA\right)\Lambda^{2N+1}\right)^\frac{1}{N+1}\sim
      \barA^{1+\frac{1}{N+1}}\Lambda^{2-\frac{1}{N+1}}\,.
  \end{equation}
It is easy to see that $\barA$ is stabilized near the origin of the moduli space thus justifying the naive analysis based on the tree level superpotential in terms of \ac{IR} degrees of freedom. 
Of course, in this model one does not have to rely on the semiclassical analysis we just performed. Indeed, the description of the theory in terms of \ac{IR} degrees of freedom is valid everywhere on the moduli space of the deformed theory, and analysis of the full superpotential (given by the sum of (\ref{eq:SP-dyn}) and (\ref{eq:SP-deform})) would yield the same result~\footnote{One must remember that while the K\"ahler potential of mesons is, in principle, calculable it is far from canonical at large $\bar A$.}. However, the semiclassical analysis is often more intuitive and, as we shall see in section \ref{sec:SON}, in some models it is the only tool at our disposal.

So far we have illustrated dynamical generation of the mass gap in models where the chiral symmetry group is $\SU{M}$ with $M=2F=2N+4$ even.  This restriction is a consequence of the fact that the fundamental of $\SP{2N}$ has an even dimension. However, it is easy to generalize this construction to models with odd $M$ \cite{Razamat:2020kyf}. Indeed, one can simply start with the same \sconf $\SP{2N}$ sector but choose the chiral group $G=\SU{M}=\SU{2F-1}$ to be a subgroup of the maximal chiral symmetry. Under $\SU{M}$ the meson $\cA$  decomposes into an antisymmetric $\cA$ and a fundamental $\cQ$. Given this choice of chiral symmetry the \ac{IR} matter content of the model is given in Table \ref{tab:SPNodd}.
 \begin{table}[h!] \centering
      \begin{tabular}{cccc}
  \toprule & \SP{2N} & $\SU{2N+3}$ & $\U{1}_R$\\ 
  \midrule $\cA$ & $\boldsymbol{1}$ & $\ytableaushort{\\}$ & $\frac{2}{N+2}\vphantom{\frac{2^2}{N+2}}$ \\
  $\cQ$ & $\boldsymbol{1}$ & $\ytableaushort{~}$ & $\frac{2}{N+2}\vphantom{\frac{2^{2^2}}{N+2}}$ \\
  $\barA$ & $\boldsymbol{1}$ & $\overline{\ytableaushort{\\}}$ & $\frac{2N+2}{N+2}\vphantom{\frac{2^{2^2}}{N+2}}$ \\
  $\barQ$ & $\boldsymbol{1}$ & $\overline{\ytableaushort{~}}$ & $\frac{2N+2}{N+2}\vphantom{\frac{2^{2^2}}{N+2}}$ \\
  \bottomrule
    \end{tabular}
    \caption{\ac{IR} content of the odd $M$ model, $M=2N+3$.}
    \label{tab:SPNodd}
  \end{table}
Since our mass gap analysis did not rely on the dynamics of the $\SU{2N+4}$ sector~\footnote{Recall that aside from requiring a cancellation of cubic anomalies we treat $G$ sector of the model as a global symmetry.}, the chirally symmetric vacuum with mass gap will exist as long as we include the tree level superpotential (\ref{eq:SP-deform}), now written in terms of $\SU{2N+3}$ degrees of freedom. Note that while the tree level superpotential must result in a maximal rank mass matrix in the \ac{IR} it does not have to respect the maximal global $\SU{2N+4}$ symmetry.

\subsection{A massless composite generation}
\label{sec:composite-gen}
In the previous subsection, we have mentioned that the choice of spectators is not unique. Rather than choosing them in representations of $\SU{2N+4}$ conjugate to those of mesons, we could choose, for example, the spectators transforming in representations conjugate to those of quark superfields $Q$. In this case, while the full theory is chiral, the \ac{UV} matter content from the point of view of the $\SU{2N+4}$ sector is non-chiral. Once the theory confines, the low energy degrees of freedom contain $\SP{2N}$ composites which transform in an antisymmetric representation of the $\SU{2N+4}$ symmetry. Thus a non-chiral $\SU{2N+4}$ sector acquires in the \ac{IR} a massles chiral generation containing an antisymmetric tensor and $2N$ antifundamentals. This theory may further be complemented by superpotential interactions between the $\SP{N}$ moduli and spectators. Consider for example an $\SP{2}\times \SU{6}$ model with matter given in \eqref{tab:SPN}. If we choose $G=\SU{3}\times \SU{2}\times \U{1}\subset \SU{5}\subset \SU{6}$ with a standard decomposition of GUT fields under the \ac{SM}, add two more spectator generations charged under the \ac{SM} and include all the superpotential terms allowed by symmetries we will arrive at the composite supersymmetric model of Nelson and Strassler \cite{Nelson:1996km}.

\subsection{Different embeddings of $G$}
\label{sec:otherembed}
We conclude the discussion of chirality flows in \sconf \SP{2N} models by noting that one can construct new models by simply choosing different embeddings of the chiral symmetry group $G$ into the maximal global symmetry of the \sconf sector. Let's briefly look at some examples. For our first example, we consider the model studied in \cite{Razamat:2020kyf} with $H=\SP{2}$ and $G=\SU{3}\times\SU{2}\times \U{1}\subset \SU{5}\subset \SU{6}$. Once again, the tree level superpotential must be the most general one consistent with $G$ but does not need to respect the full $\SU{6}$ global symmetry of the s-confining sector. A somewhat more elaborate example can be found by considering $N=3$ case, i.e. an s-confining $\SP{6}$ model with $5$ flavors and $\SU{10}$ global symmetry. We will take  $G=\SU{5}$ and embed it into $\SU{10}$ global symmetry so that $10$ quark superfields transform in an antisymmetric representation of $\SU{5}$. The mesons $M\sim Q^2$ then transform as $\bf{45}$ of $\SU{5}$. We now add the spectators in the $\overline{\bf{45}}$ representation of $\SU{5}$. The analysis of strong $\SP{6}$ dynamics remains unchanged and the model develops a mass gap in the \ac{IR}. In our final example we start with the same $\SP{6}$ \sconf sector and choose $G=\SU{3}$ embedding it into $\SU{10}$ in such a way that $10$ quark superfields transform in a $3$-index symmetric representation of $\SU{3}$. The $\SP{6}$ mesons decompose as $\rep{10}\oplus\rep{35}$ of $\SU{3}$. Adding spectators in $\overline{\rep{10}}\oplus\overline{\rep{35}}$ representations as well as the most general superpotential results in a mass gap appearing in the \ac{IR}. 
Our discussion so far suggests that, in addition to generating mass gaps or composite chiral matter, chirality flows may lead to more general results. Indeed, in the following sections, we will see examples of models where both \ac{UV} and \ac{IR} physics is chiral yet the chiral structure of the theory changes in the course of \ac{RG} flow. Our examples will include models based on different \sconf sectors but even within specific \sconf dynamics we will have the freedom to construct different models of chirality flow by using two different tools: an ability to choose different representations of spectators introduced to cancel anomalies and use of different embeddings of $G$ into the maximal global symmetry of the s-confining sector. 
	
	\section{The role of s-confinement: an $\SO{N}$ example} \label{sec:SO}
\label{sec:SON}

  In the previous section, we analyzed the dynamics of models
  where mass gap is generated in the IR despite the matter content being chiral in
  the UV. Following \cite{Razamat:2020kyf} our examples were based on $s$-confining $\SP{2N}$ theories and the choices of chiral matter representations were dictated by embedding of the chiral symmetry in the maximal global symmetry of the $s$-confining model. The simplest and most illuminating embedding generated a mass gap in models with matter transforming in an antisymmetric representation of the chiral $\SU{N}$ symmetry. This was a consequence of the fact that the composites of $\SP{2N}$ models transform as antisymmetrics of global symmetries. It is then natural to expect that chiral matter may be gapped in models where the composites of the confining sector transform in symmetric representations of the global symmetry. To that end, the authors of \cite{Tong:2021phe} argued that a mass gap in theories with symmetric chiral matter can be generated by deformations of confining  $\SO{N}$ sector with $F=N-4$ chiral superfields in a vector representation. It is known \cite{Intriligator:1995id} that this class of models exhibits two phases: a phase with dynamically generated runaway superpotential and a no-superpotential phase where quantum moduli space coincides with the classical one and extends to the origin. It was argued in \cite{Tong:2021phe} that an appropriate deformation of these models generates a mass gap in the no-superpotential phase. Unfortunately, this class of $\SO{N}$ models is not $s$-confining \cite{Csaki:1996zb} and the phase with chirally symmetric vacuum is quite fragile. We will argue
  here that the deformations necessary to generate the gap destroy the
  chirally symmetric vacuum. Fortunately, as we will show in section \ref{sec:sun} constructions of
  gapped symmetric fermion models are still possible albeit they are more
  cumbersome than hoped for in \cite{Tong:2021phe}.

  \subsection{An $\SU{F} \times \SO{F+4}$ model}
  We begin the analysis by reviewing the dynamics of $\SO{N}$ theories with $F=N-4$ flavors
  \cite{Intriligator:1995id}. The quantum numbers of the matter fields under
  the gauge $\SO{N}$ and global $\SU{N-4}$ symmetries are given in Table \Cref{tab:SON}.
  \begin{table}[h!] 
    \centering 
    \begin{tabular}{cccc} 
      \toprule & \SO{N} & $\SU{F}$ & $\U{1}_R$\\ 
      \midrule $Q$ & $\ytableaushort{~}$ & $\ytableaushort{~}$ &
        $\frac{F-N+2}{F}$  \\ 
      \midrule $\cS$ & $\boldsymbol{1}$ & $\ytableaushort{~~}$ &
        $\frac{2(F-N+2)}{F}$ \\ $\overline{S}$ & $\boldsymbol{1}$ &
      $\overline{\ytableaushort{~~}}$ & $\frac{2(N-2)}{F}\vphantom{\frac{2(N-2)^{2^2}}{F}}$\\ 
      \bottomrule 
    \end{tabular}
    \caption{Field content of the $\SO{N}$ model with $F$ flavors.}
    \label{tab:SON}
  \end{table}\\
  The one-loop beta function of $\SO{N}$ theory, for $N > 4$ is 
  \begin{equation}
  	b = 3(N-2) - F\;.
  \end{equation}
  The classical moduli space can be parameterized in terms of quark \acp{VEV} or gauge invariant mesons $\cS_{ij}=Q_iQ_j$. At a generic point on this moduli space the gauge group is broken to a pure \ac{SYM} $\SO{4} \sim \SU{2}_L \times \SU{2}_R$. Further, in the \ac{IR} $\SU{2}_L \times \SU{2}_R$ group confines, generating the gaugino condensate superpotential. Since the dynamical scale of the low energy physics depends on the moduli, this results in the superpotential for $\SO{N}$ fields which, in terms of mesons, takes the form
  \begin{align} 
    W_{\text{dyn}} &= 2\langle \lambda\lambda\rangle_L + 2\langle
    \lambda\lambda\rangle_R \\ &= \frac{1}{2}(\epsilon_L +
    \epsilon_R)\left(\frac{16\Lambda^{2(N-1)}}{\text{det}
    \cS}\right)^{1/2}\;,
  \end{align} 
  where $\epsilon_{L,R} = \pm 1$. As explained in \cite{Intriligator:1995id} the theory has two phases. When
  $\epsilon_L\epsilon_R=1$ the two contributions to the superpotential add up
  constructively and the classical moduli space is lifted, resulting in a phase
  without a stable ground state. When $\epsilon_L\epsilon_R=-1$ the two
  contributions to the superpotential cancel\footnote{A pure SYM $\SO{4}$ theory
    is described by two dynamical scales, $\Lambda_L$ and $\Lambda_R$ which need
    not be equal. However, in our case the dynamical scales of the low energy
    gauge groups are determined uniquely (up to a sign) by the dynamical scale of UV
    physics and mesons VEV, thus ensuring the cancellation of the two terms in the
    superpotential.}\!,
resulting in a smooth quantum moduli space with an unbroken $\SO{N}$ chiral
  global symmetry at the origin.  
  
  Let us now deform the theory by including superfields $\overline{S}$ transforming in conjugate symmetric reprsentation of the chiral $\SU{F}=\SU{N-4}$ symmetry\footnote{It is easy to see that the matter content is anomaly free under the full $\SO{N}\times \SU{N-4}$ symmetry.}. Since the low energy matter content is vector-like we can include a tree level superpotential which appears as a mass term in the \ac{IR}. The full low energy
  superpotential takes the form
  \begin{align} 
    W = \frac{1}{2}(\epsilon_L +
    \epsilon_R)\left(\frac{16\Lambda^{2(N-1)}}{\text{det}\cS}\right)^{1/2} +
    \Lambda\cS\overline{S}\;. 
  \end{align} 
  A naive analysis of the no-superpotential branch suggests that our deformation generates a mass gap.    However, the absence of $s$-confinement and the presence of a second, runaway,  phase in $\SO{N}$ models implies that, in contrast to the
  theories we discussed earlier, the chirally symmetric vacuum is unstable under any deformation.
  For example, an explicit mass term, $mQ^2$, lifts the classical moduli space while remaining consistent
  with an existence of the chirally symmetric vacuum at the origin. Yet, as
  argued in \cite{Intriligator:1995id}, at the quantum level the full no-superpotential branch, including the chirally symmetric vacuum, is lifted.
	
  To better understand the fate of the chirally symmetric phase we will study the non-perturbative dynamics in a weakly coupled regime. We note that the deformed theory possesses a new classical flat direction parameterized by $\overline{S}$. When $\overline{S}\gg\Lambda_{\SO{N}}$ the physics is weakly coupled and the K\"ahler potential is nearly canonical in terms of the quark superfields. Furthermore, along this flat direction the quarks become massive, suggesting
  that argument of \cite{Intriligator:1995id} for the disappearance of the chirally symmetric vacuum should apply. The dynamical nature of $\overline{S}$ allows us to perform a more
  detailed analysis. At large $\overline{S}$ the quarks must be integrated out, and the
  low energy physics is described by a pure $\SO{N}$ SYM theory with the dynamical scale given by 
  \begin{equation}
    \Lambda_{L}^{3(N-2)}=\det{\overline{S}}\Lambda^{2N-2}\,. 
  \end{equation} 
The low energy physics then generates the dynamical superpotential
  \begin{equation}
    W=\Lambda_L^3=\left(\det\overline{S}\right)^{\frac{1}{N-2}}\Lambda^{2+\frac{2}{N-2}}\,.
  \end{equation} 
  One can see that this superpotential leads to runaway behavior for $\overline{S}$.
  While our derivation is only valid at large values of $\overline{S}$, holomorphy
  suggests that in the absence of a singularity in the
   K\"ahler potential the SUSY vacuum at the origin must be
  lifted.

	\section{Chirality flows and \SU{N} dynamics} \label{sec:SU}
\label{sec:sun}	
  \subsection{\sconf SQCD}
	
  We begin by briefly reviewing an \sconf SQCD model with $F=N+1$ flavors. The theory
  has an $\SU{N+1}_\mathrm{L}\times \SU{N+1}_\mathrm{R}\times \U{1}_B\times
  \U{1}_R$ anomaly-free global symmetry and the matter charges under
  gauge and global symmetries are given in the top part of \Cref{tab:SQCD1}.
  \begin{table}[h!] 
    \centering 
    \begin{tabular}{cccccc} 
      \toprule & \SU{N} & $\SU{F}_\mathrm{L}$ & $\SU{F}_\mathrm{R}$ & $\U{1}_B$
               & $\U{1}_R$\\
      \midrule $Q$ & $\ytableaushort{~}$ & $\ytableaushort{~}$ & $\rep{1}$ &
               $1$ & $\frac{1}{N+1}$\\
      $\overline{Q}$ & $\overline{\ytableaushort{~}}$ & $\rep{1}$ &
               $\overline{\ytableaushort{~}}$  & $-1$ & $\frac{1}{N+1}\vphantom{\frac{2^{2^2}}{N+2}}$\\
      \midrule $\cM$ & $\rep{1}$ & $\ytableaushort{~}$ &
               $\overline{\ytableaushort{~}}$  & $0$ & $\frac{2}{N+1}$\\ 
      $\cB$ & $\rep{1}$ & $\overline{\ytableaushort{~}}$ & $\rep{1}$ & $N$ &
               $\frac{N}{N+1}\vphantom{\frac{2^{2^2}}{N+2}}$\\ 
      $\barcB$ & $\rep{1}$ & $\rep{1}$ & $\ytableaushort{~}$  & $N$ &
               $\frac{N}{N+1}$\\ 
      \midrule $S$ & $\rep{1}$ & $\overline{\ytableaushort{~}}$ &
               $\ytableaushort{~}$  & $0$ & $\frac{2N}{N+1}$\\ 
      $\overline{B}$ & $\rep{1}$ & $\ytableaushort{~}$ & $\rep{1}$ & $-N$ &
               $\frac{N+2}{N+1}\vphantom{\frac{2^{2^2}}{N+2}}$\\ 
      $B$ & $\rep{1}$ & $\rep{1}$ & $\overline{\ytableaushort{~}}$  & $-N$ &
               $\frac{N+2}{N+1}\vphantom{\frac{2^{2^2}}{N+2}}$\\
      \bottomrule
    \end{tabular}
    \caption{Field content of \sconf SQCD model with $F=N+1$ flavors. The top
      portion of the table shows the elementary \SU{N} charged fields. The
      middle section of the table shows the confined degrees of freedom that are
      weakly coupled in the IR and near the origin of the moduli space. The
      bottom portion of the table shows the quantum numbers of the spectator
      fields needed to cancel flavor symmetry anomalies and generate mass gap
      for chiral fermions in the IR.}
    \label{tab:SQCD1}
  \end{table}
  The existence of a large chiral symmetry will allow us to construct a variety of models exhibiting chirality changing RG flows by considering different embeddings of $G$ in the maximal global symmetry group of the \sconf sector.
  
  In the absence of the superpotential, the model possesses a large moduli space of classical flat directions. These flat directions can be parameterized in terms of gauge invariant composites, $\cM\sim Q\overline{Q}$, $\cB\sim Q^N$, and $\barcB\sim \overline{Q}^N$, whose quantum numbers are presented in the middle section of \Cref{tab:SQCD1}. Classically the moduli \acp{VEV} satisfy a set of identities
\begin{equation}
    \cM_{ij}\barcB_{j}=\cB_{j}M_{ji}=
      \cB_{j}\barcB_i=0\,. 
  \end{equation}
	
It is well known that  this model $s$-confines and the low energy physics is described in terms of mesons, baryons, and anti-baryons with dynamically generated superpotential which implements the classical constraints at the quantum level
  \begin{equation} \label{eq:sconfinedW}
    \mathscr{W}=\frac{1}{\Lambda^{2N-1}}\left( \cB\cM\barcB-\det \cM\right)\,.
  \end{equation} 
  In the IR, mesons and baryons are weakly coupled and have a nearly canonical
  K\"ahler potential. Thus it is convenient to rescale the moduli by
  absorbing appropriate powers of the dynamical scale into the definition of the
  moduli so that $\cM$, $\cB$, and $\overline{\cB}$ have mass dimension one.
	
  In the following subsections, we will consider several illustrative embeddings of a chiral group $G$ in the maximal global symmetry of the \sconf sQCD where a chirally symmetric vacuum is preserved while a mass gap is developed.
	
  \subsection{$G=\SU{N+1}_\mathrm{L}\times\SU{N+1}_\mathrm{R}$}
	
  As our first example, we choose\footnote{The following analysis remains
  unchanged if we include a $\U1_B$ factor in the definition of $G$.}
  $G=\SU{N+1}_L\times\SU{N+1}_R$. As discussed earlier, the low energy content
  of $G$ is given by mesons $\cM$, baryons $\cB$, and anti-baryons $\barcB$
  transforming as $(\ytableaushort{~}\,, \overline{\ytableaushort{~}})$,
  $(\overline{\ytableaushort{~}}\,,1)$, and $(1,\ytableaushort{~})$,
  respectively. Since our goal is to deform this model in such a way that $G^3$ anomalies vanish while the low energy matter content is vector-like, we introduce a set of spectators, 
  $\overline{M}$, $B$, and $\overline{B}$ in representations conjugate to those
  of $\cM$, $\cB$, and $\barcB$. For completeness, the quantum numbers of the
  spectator fields are displayed in the bottom portion of \Cref{tab:SQCD1}.
	
  The inclusion of the spectator fields in the theory allows a tree-level
  superpotential consistent with the full $H\times G$ symmetry,
  \begin{equation} \label{eq:Wspectators}
    \mathscr{W}_\mathrm{tree}=\overline{M}\cM+\overline{B}Q^N+B {\overline{Q}}^N\sim\overline{M}\cM+\overline{B}\cB+B \barcB \,.
  \end{equation} 
  Repeating the analysis of section \ref{sec:SP} far along the meson branch of the moduli space we obtain the low energy superpotential for mesons
  \begin{equation} \label{eq:SUN-stabilize}
    \mathscr{W}=\Lambda_L^3=\left(\det\barM
    \Lambda^{2N-1}\right)^{1/N}\,,
  \end{equation} 
  where we used the scale matching relation
  \begin{equation} \label{eq:SUN-match} 
    \Lambda_L^{3N}=\det\barM \Lambda^{2N-1}\,.
  \end{equation}

 We see that this superpotential stabilizes the spectator mesons at the origin of the moduli space. The analysis of baryonic directions is more complicated due to the non-renormalizability of the superpotential terms involving the baryons in the \ac{IR}. Nevertheless, an analysis of the full superpotential shows that the baryons are also stabilized at the origin. Having established the absence of runaway directions on the moduli space we conclude that this model develops a mass gap.

  \subsection{ $G=\SU{N+1}_D$ with symmetric and antisymmetric}
  \label{sec:AplusS}
  Our next example involves the identification of $G$ with an $\SU{N+1}_D$
  diagonal subgroup of $\SU{N+1}_L\times\SU{N+1}_R$. However, if this diagonal
  subgroup is generated by $T_D=T_L+T_R$ the matter fields transform in 
  non-chiral representations of $\SU{N+1}_D$ and thus this case is not of interest to us.
  Instead, we will consider $\SU{N}_D$ generated by $T_D=T_L-T_R$. The easiest
  way to do so is to assign $\overline{Q}$ to a fundamental rather than antifundamental representation of $\SU{N+1}_R$. With this charge assignment, the mesons $\cM$
  transform as a sum of symmetric and antisymmetric
  representations of $G$ while both baryons and anti-baryons transform in 
  anti-fundamental representation. This implies that the spectator field $\barM$ decomposes as $\overline{A}$ and $\overline{S}$, while both $B$ and $\overline{B}$ transform as fundamentals of $\SU{N+1}_D$. The matter content of this model is given in \Cref{table:AplusS}. The deformation superpotential 
  (\ref{eq:Wspectators}) takes the form
  \begin{equation}
  \label{eq:AplusS}
   W=\bar A\cA+\barS\cS+\overline{B}\cB+B \barcB
  \end{equation}
The non-perturbative dynamics of the model remains unchanged and vacuum is found at the origin of the moduli space.
  \begin{table}[h!] 
  	\centering 
	\begin{tabular}{cccc} 
		\toprule & \SU{N} & $\SU{N+1}_D$ & $U(1)_B$\\
		\midrule $Q_i$ & $\ytableaushort{~}$ & $\ytableaushort{~}$ & $1$ \\ 
		$\overline{Q}_i$ & $\overline{\ytableaushort{~}}$ & $\ytableaushort{~}$ & $-1$\\
	\midrule $\cM = \cA + \cS$ & $\rep{1}$ & $\ytableaushort{~~} \oplus \ytableaushort{\\} $ & $0$\\
	$\cB$ &  $\rep{1}$ & $\overline{\ytableaushort{~}}$ & $N$\\
	$\barcB$ &  $\rep{1}$ & $\overline{\ytableaushort{~}}$ & $-N$\\
	\midrule $\barM = \barA + \barS$ & $\rep{1}$ & $\overline{\ytableaushort{~~}} \oplus \overline{\ytableaushort{\\}} $ & $0$\\
	$\overline{B}$ &  $\rep{1}$ & ${\ytableaushort{~}}$ & $-N$\\
	$B$ &  $\rep{1}$ & ${\ytableaushort{~}}$ & $N$\\
	\bottomrule
	\end{tabular} 
	\caption{Field content of the $\SU{N}$ model with $\SU{F}_D$ flavor symmetry}
	\label{table:AplusS}
\end{table}
	
 Simply by choosing a different chiral symmetry group $G$ and selecting a desirable embedding of this group in the maximal global symmetry of \sconf $\SU{N}$ we have constructed a model with one chiral symmetric and one chiral antisymmetric representation in the \ac{UV} which is fully gapped in the \ac{IR}.

  \subsection{Antisymmetric $\leftrightarrow$ Symmetric Flows} 
  \label{sec:AtoS}
  The early studies of the chirality flows \cite{Nelson:1996km,Strassler:1995ia} aimed at generating composite chiral matter in the \ac{IR} while the recent work \cite{Razamat:2020kyf,Tong:2017oea} was driven by an interest in generating mass gaps in chirally symmetric vacua. In this section, we will illustrate that these two types of models are simply extreme examples of a more general class of chiral theories where the chirality structure changes under the  \ac{RG} flow. Indeed, we have already used the fact that the choice of spectators necessary to cancel $G^3$ anomalies is not unique. To generate the mass gap we chose the spectators in representations of $G$ conjugate to the representations of the composites of the strong dynamics. On the other hand, to generate composite chiral matter in the \ac{IR} we chose the spectators in the representations of $G$ conjugate to representations of the elementary superfields. But one can mix and match. For example, in the model of section \ref{sec:AplusS} we can replace $\barA$ with $N-4$ spectators $\bar q$ transforming as antifundamentals of $\SU{N+1}_D$. In this case, the \ac{UV} model contains a chiral symmetric representation of $\SU{N+1}_D$ and $N-4$ vector-like flavors (with all antiquarks of $\SU{N+1}_D$ being spectators and all quarks charged under $\SU{N}$).
With this choice of $G$ and the spectator fields the most general tree level superpotential is
  \begin{equation}
  \label{eq:AtoS}
   W=y \cA\bar q \bar q +\barS\cS+\overline{B}\cB+B \barcB\,,
  \end{equation}
  where we have explicitly included the Yukawa coupling $y$ in the first term. We note in passing that $y$ is naturally small since it arises from a non-renormalizable term in the \ac{UV} description.
  Analyzing the non-perturbative dynamics of this model we find that in the \ac{IR} the composite $\cS$ and the spectator $\overline{S}$ pick up a mass and decouple from the low energy physics while the massless matter content contains a single antisymmetric generation. Thus we constructed a more general model of chirality flow where non-perturbative dynamics modifies the chiral structure of the theory in \ac{IR} instead of simply adding or removing a chiral generation. A reverse flow, from an antisymmetric generation in the \ac{UV} to a symmetric generation in the \ac{IR}, is equally easy to achieve. 
  
  \subsection{Gapping symmetric matter}
The results of Section {\ref{sec:AtoS} suggest a model-building trick that allows one to gap the symmetric $S$ of the chiral $G=\SU{F}$ symmetry, even if the required model is somewhat baroque. To that end, one needs two \sconf sectors, both with fields charged under $G$. The first \sconf sector is based on an $\SU{N}$, $N=F-1$, gauge group whose composites transform as $\cA$ and $\cS$ of $G$, while the second sector is based on $\SP{2M}$, $2M=F-4$, group whose composites transform as $\barcA$. The matter content is given in \Cref{tab:gappedS}.
  
\begin{table}[h!] 
	\centering 
	\begin{tabular}{cccc} 
		\toprule & \SU{N} & $\SU{F}_D$ & $\SP{2M}$\\
		\midrule $Q_i$ & $\ytableaushort{~}$ & $\overline{\ytableaushort{~}}$ & $\rep{1}$\\ 
		$\overline{Q}_i$ & $\overline{\ytableaushort{~}}$ & $\overline{\ytableaushort{~}}$ & $\rep{1}$\\
		$q$ & $\rep{1}$ & $\ytableaushort{~}$ & $\ytableaushort{~}$\\
		\midrule
		$\barcS \oplus \barcA$ & $\rep{1}$ & $\overline{\ytableaushort{~~}} \oplus \overline{\ytableaushort{\\}}$ & $\rep{1}$\\
		$\cB$ & $\rep{1}$ & $\ytableaushort{~}$ & $\rep{1}$\\
		$\barcB$ & $\rep{1}$ & $\ytableaushort{~}$ & $\rep{1}$\\
		$\cA$ & $\rep{1}$ & $\ytableaushort{\\}$ & $\rep{1}$\\
		\midrule
		$S$ & $\rep{1}$ & $\ytableaushort{~~}$ & $\rep{1}$\\
		$B$ &  $\rep{1}$ & $\overline{\ytableaushort{~}}$ & $\rep{1}$\\
		$\overline{B}$ &  $\rep{1}$ & $\overline{\ytableaushort{~}}$ & $\rep{1}$\\
		\bottomrule\\
	\end{tabular} 
	\caption{Field content of the gapped symmetric model. The top section shows elementary fields of the model charged under one of the \sconf sectors, the middle section shows the composites of strong dynamics, and the bottom section shows the spectators charged only under the chiral $G=\SU{F}$ symmetry}
	\label{tab:gappedS}
\end{table}
The tree level superpotential in terms of composites and the spectators  is given by
\begin{equation}
 W_\mathrm{tree}=\barcS S+\barcA \cA +B\barcB+\overline{B}\cB\,.
\end{equation}
A careful analysis of dynamics  in regions where either $\cA$ or $\barcA$ is large establishes that the model develops a mass gap in the \ac{IR}.

  \section{Summary} \label{sec:conc}

  In this paper we have conducted a detailed investigation into the non-perturbative dynamics underlying chirality flows in strongly interacting \ac{SUSY} gauge theories. Our results suggest that chirally symmetric vacua are stable under required deformations if and only if the strongly interacting sector satisfies \sconfinement criteria \cite{Seiberg:1994bz,Csaki:1996zb}. We analyzed an example of an $\SO{N}$ model which, in the absence of deformations, possesses a phase with chirally symmetric vacuum and showed that in the deformed theory this vacuum is destabilized by the interplay between the non-perturbative dynamics and the tree level superpotential.\\
We also developed model building tools that allow the construction of various models exhibiting chirality flows, including cases where chiral matter transforming either in symmetric or antisymmetric representations can be gapped. Beyond looking at dynamically generated mass gaps in chiral models, we have presented a more universal approach to the study of dynamics underlying chirality flows. Although our study focused on theories with no tree-level superpotential in the UV, it would be interesting to explore generation flow in theories with tree-level superpotentials, as discussed in \cite{Csaki:1998fm}.

\section{Acknowledgments}
We thank Sa\'ul~Ramos--S\'anchez and Michael Ratz for many useful discussions. Y.S. is grateful to Technion and ICTP-SAIFR where parts of this work were completed. The work of S.S. and Y.S. was supported in part by the US National Foundation under grant PHY-2210283.  The work of S.S. was supported in part by UC-MEXUS-CONACyT grant No. CN-20-38. M.W. was  supported in part by the Israel Science Foundation
(Grant No.~751/19), by the NSF-BSF (Grant No.~2020-785) and by a Zuckerman Fellowship.
  \bibliographystyle{NewArXiv} \bibliography{SSW_final}
  \end{document}